\global\def\draftcontrol{0}

   \def\versionno{ shear viscosity at phase transitions }
\catcode`\@=11

\expandafter\ifx\csname draftcontrol\endcsname\relax\global\def\draftcontrol{0}
\fi

{\count255=\time\divide\count255 by 60
\xdef\hourmin{\number\count255}
\multiply\count255 by-60\advance\count255 by\time
\xdef\hourmin{\hourmin:\ifnum\count255<10 0\fi\the\count255}}
\def\draftdate{\number\month/\number\day/\number\year\ \ \ \hourmin }

\newcommand\makepapertitle{\par
  \begingroup
    \renewcommand\thefootnote{\@fnsymbol\c@footnote}%
    \def\@makefnmark{\rlap{\@textsuperscript{\normalfont\@thefnmark}}}%
    \long\def\@makefntext##1{\parindent 1em\noindent
            \hb@xt@1.8em{%
                \hss\@textsuperscript{\normalfont\@thefnmark}}##1}%
     \newpage
     \global\@topnum\z@   % Prevents figures from going at top of page.
     \@makepapertitle
     \thispagestyle{empty}\@thanks
  \endgroup
  \setcounter{footnote}{0}%
  \global\let\thanks\relax
  \global\let\makepapertitle\relax
  \global\let\@makepapertitle\relax
  \global\let\@thanks\@empty
  \global\let\@author\@empty
  \global\let\@date\@empty
  \global\let\@title\@empty
  \global\let\title\relax
  \global\let\author\relax
  \global\let\date\relax
  \global\let\and\relax
  \def\version{\let\version\@version\@gobble}
}
\def\@makepapertitle{%
  \newpage
   \ifnum\draftcontrol=1 {}
   \version\versionno
   \vskip 3em%
   \else
   \hfill\hbox to 3cm {\parbox{4cm}{\@pubnum}\hss}%
   \vskip 3em%
   \fi
   \begin{center}%
   \let \footnote \thanks
     {\LARGE {\@title}}%
     \vskip 1.5em%
     {\normalsize%\large
       \lineskip .5em%
       \begin{tabular}[t]{c}%
         \@author
       \end{tabular}\par}%
     \vskip 1.5em%
     {\@bstract}%
     \end{center}%
     \vskip 1.5em
     \@date%
   \par
}

\gdef\@pubnum{}
\def\pubnum#1{%
  \gdef\@pubnum{#1}}

\gdef\@bstract{}
\def\Abstract#1{%
  \gdef\@bstract{%
   \parbox{\textwidth-0pc}{%
   \centerline{\bf Abstract}\penalty1000%
\kern.2cm%
\noindent%\abstractfont \baselineskip=12pt
\renewcommand\baselinestretch{1.0}%
{#1}}}
}

\def\ps@paper{\let\@mkboth\@gobbletwo%
     \ifnum\draftcontrol=1
    \def\@oddfoot{\hbox to \textwidth{\tiny \versionno \hfil\tiny\draftdate}%
    \hskip -\textwidth \hbox to \textwidth{\hfil\rm\thepage\hfil}}%
     \else\def\@oddfoot{\hbox to \textwidth{\hfil\rm\thepage\hfil}}
     \fi
     \let\@evenfoot\@oddfoot
}
\def\body{\clearpage
          \pagestyle{paper}
    }
\def\@version#1{\ifnum\draftcontrol=1
\typeout{}\typeout{#1}\typeout{}
\vskip3mm\centerline{\hbox{\fbox{\normalsize{\tt DRAFT -- #1 -- }
                   {\draftdate}}}}\vskip3mm
\fi}
\let\version\@version
\long\def\eqlabel#1{\ifnum\draftcontrol=1
                    \tag@false  % there are some problems with multline without this
                    \tag*{(\theequation) \hbox to -0.2cm{\hspace{0cm}\small{#1}\hss}}
                    \refstepcounter{equation}
                    \edef\@currentlabel{\theequation}
                    \ltx@label{#1}          % use old LaTeX \label instead of new definition
                    \else
                    \label{#1}
                    \fi
                    }
\let\st@bibitem\@bibitem
\let\st@lbibitem\@lbibitem
\ifnum\draftcontrol=1
  \def\@bibitem#1{%
    \st@bibitem{#1}\a@@label{#1}\ignorespaces}
  \def\@lbibitem[#1]#2{%
    \st@lbibitem[#1]{#2}\a@@label{#2}\ignorespaces}
  \def\a@@label#1{%
    \gdef\a@lab{\smash{\normalfont\small#1}}
    \ifvmode
      \if@inlabel
        \global\setbox\@labels\hbox{%
          \llap{\a@lab\let\a@lab\relax
                \kern\@totalleftmargin\kern\marginparsep}%
          \box\@labels}%
      \fi
    \fi}
\fi
\documentclass[12pt,letterpaper]{article}
\usepackage{amsmath,amssymb,array,calc,epsfig,rotating,bm}
\usepackage[sort]{cite}
\usepackage{graphicx}
\usepackage{psfrag,verbatim}

\ifnum\draftcontrol=1
\fi
\renewcommand\baselinestretch{1.25}
\renewcommand\section{\@startsection {section}{1}{\z@}%
                                   {-3.5ex \@plus -1ex \@minus -.2ex}%
                                   {2.3ex \@plus.2ex}%
                                   {\normalfont\large\bfseries}}
\renewcommand\subsection{\@startsection{subsection}{2}{\z@}%
                                   {-3.25ex\@plus -1ex \@minus -.2ex}%
                                   {1.5ex \@plus .2ex}%
                                   {\normalfont\normalsize\bfseries}}
\renewcommand\subsubsection{\@startsection{subsubsection}{3}{\z@}%
                                   {-3.25ex\@plus -1ex \@minus -.2ex}%
                                   {1.5ex \@plus .2ex}%
                                   {\normalfont\normalsize\it}}
\renewcommand\paragraph{\@startsection{paragraph}{4}{\z@}%
                                   {-3.25ex\@plus -1ex \@minus -.2ex}%
                                   {1.5ex \@plus .2ex}%
                                   {\normalfont\normalsize\bf}}

\numberwithin{equation}{section}

\def\revise#1       {\raisebox{-0em}{\rule{3pt}{1em}}%
                     \marginpar{\raisebox{.5em}{\vrule width3pt\
                     \vrule width0pt height 0pt depth0.5em
                     \hbox to 0cm{\hspace{0cm}{%
                     \parbox[t]{4em}{\raggedright\footnotesize{#1}}}\hss}}}}

\newcommand\nxt[1]  {\\\fnxt#1}

\def\calb         {{\cal B}}
\def\calc         {{\cal C}}

\def\cale         {{\cal E}}

\def\calg         {{\cal G}}

\def\calk         {{\cal K}}
\def\call         {{\cal L}}

\def\caln         {{\cal N}}
\def\calo         {{\cal O}}

\def\del          {\partial}

\def\half{{\frac12}}

\def\sqr#1#2{{\vcenter{\vbox{\hrule height.#2pt
 \hbox{\vrule width.#2pt height#1pt \kern#1pt
 \vrule width.#2pt}\hrule height.#2pt}}}}

\def\a{\alpha}
\def\b{\beta}

\newcommand{\ww}{\mathfrak{w}}

\newcommand{\beq}{\begin{equation}}
\newcommand{\eeq}{\end{equation}}
\newcommand{\beqa}{\begin{eqnarray}}
\newcommand{\eeqa}{\end{eqnarray}}
\newcommand{\beqar}{\begin{eqnarray*}}
\newcommand{\eeqar}{\end{eqnarray*}}

\newcommand{\labell}[1]{\label{#1}}
\renewcommand{\eqref}[1]{(\ref{#1})}

\newcommand{\ie}{{\it i.e.,}\ }

\newcommand{\ka}{\bm{k}}

\def\a{\alpha}
\def\w{\omega}

\def\om{\Omega}

\def\p{\varphi}
\def\pp{\phi}
\def\r{\rho}

\def\l{\lambda}
\def\lgb{\lambda_{GB}}

\catcode`\@=12

\begin{document}

%%%
%%%%%% text starts here
%%%%%%%%%

\title{\bf Viscosity Bound and Causality \\in  Superfluid Plasma }
\pubnum{UWO-TH-10/04 DAMTP-2010-52 MIFPA-10-32}

\date\today

\author{
Alex Buchel$ ^{1,2}$ and Sera Cremonini$ ^{\,\clubsuit,\spadesuit}$ \\[0.4cm]
\it $ ^1$Perimeter Institute for Theoretical Physics\\
\it Waterloo, Ontario N2L 2Y5, Canada\\[.5em]
 \it $ ^2$Department of Applied Mathematics\\
 \it University of Western Ontario\\
\it London, Ontario N6A 5B7, Canada\\[.5em]
\it $ ^\clubsuit$ Centre for Theoretical Cosmology, DAMTP, CMS,\\
\it University of Cambridge, Wilberforce Road, Cambridge, CB3 0WA, UK \\ [.5em]
\it $ ^\spadesuit$ George and Cynthia Mitchell Institute for Fundamental Physics and Astronomy\\
\it Texas A\&M University, College Station, TX 77843--4242, USA
 }

%%%%

\Abstract{It was argued by Brigante et.al \cite{rob1} that the lower bound on the
ratio of the shear viscosity to the entropy density in strongly coupled plasma
is translated into microcausality violation in the dual gravitational description.
Since transport properties of the system characterize its infrared dynamics, while the
causality of the theory is determined by its ultraviolet behavior, the viscosity bound/microcausality
link should not be applicable to theories that undergo low temperature phase transitions.
We present an explicit model of AdS/CFT correspondence that confirms this fact.
}

\makepapertitle

\body

\version\versionno
\tableofcontents

\section{Introduction}
Working in the framework of the gauge theory/string theory correspondence of Maldacena \cite{m9711,m2},
Policastro, Son and Starinets computed the ratio of the shear viscosity $\eta$ to entropy density $s$
of the $\caln=4$ $SU(N)$ supersymmetric Yang-Mills (SYM) plasma, in the planar ('t Hooft) limit
and for infinitely large 't Hooft coupling\footnote{We set $\hbar=k_B=1$.}
$\l=g_{YM}^2 N\to \infty$ \cite{pss}, finding
\begin{equation}
\frac{\eta}{s}=\frac{1}{4\pi} \, .
\eqlabel{univ}
\end{equation}
Shortly afterwards it was argued in \cite{u1} that \eqref{univ} is in fact a universal result in
all gauge theory plasma at infinite coupling that allow for a dual holographic
description\footnote{Further generalizations/proofs of the shear viscosity universality
theorem appeared in \cite{u2,u3,bu,u5,u6,u7}.}.

The holographic result \eqref{univ} is remarkable in a sense that a simple quasi-particle picture of
hydrodynamic transport suggests a quantum mechanical bound \cite{ss}
\begin{equation}
\frac{\eta}{s}\gtrsim \calo\left(1\right) \, .
\eqlabel{bound0}
\end{equation}
This fact, along with the observation that all known fluids in nature\footnote{A strongly
coupled Quark-Gluon Plasma might be a counterexample \cite{lr}.} have  larger shear viscosity
to entropy density ratios, led Kovtun, Son and Starinets (KSS) to conjecture
a bound for {\it any fluid} \cite{u2}:
\begin{equation}
\frac{\eta}{s}\ge \frac{1}{4\pi} \, .
\eqlabel{bound}
\end{equation}
It is possible to construct a phenomenological counterexample in which, by
increasing the number of species in the fluid while keeping the dynamics essentially independent of the species type,
the bound can be violated \cite{c}. Unfortunately, the particular example \cite{c} does not
have a well-defined relativistic quantum field theory completion \cite{cs}.

The first test confirming  the KSS bound, at least  for  $\caln=4$ SYM at
large  (but finite) 't Hooft coupling, was done in
\cite{bls}\footnote{For further analysis and generalizations see \cite{f1,f2,f3,f4,f5}.}.
The finite 't Hooft coupling corrections on the gauge theory side translate into
higher-derivative gravitational corrections on the string theory side of the
holographic correspondence \cite{m2}.
This model, along with  generalizations \cite{f3,f4}, describes a superconformal
gauge theory plasma --- a consistent relativistic quantum field theory --- with the same
anomaly coefficients (central charges) $c=a$ in the trace of the stress-energy tensor,
\begin{equation}
\langle T^\mu{}_\mu\rangle_{CFT} =\frac{c}{16\pi^2}
I_4-\frac{a}{16\pi^2} E_4\,. \labell{confag}
\end{equation}
Here $E_4$ and $I_4$ correspond, respectively, to the four-dimensional Euler density and the
square of the Weyl curvature:
\begin{equation}
E_4= R_{\mu\nu\rho\lambda}R^{\mu\nu\rho\lambda}-4
R_{\mu\nu}R^{\mu\nu}+R^2 \,,\qquad I_4=
R_{\mu\nu\rho\lambda}R^{\mu\nu\rho\lambda}-2  R_{\mu\nu}R^{\mu\nu}
+\frac 13R^2\,. \labell{ejdef}
\end{equation}
In \cite{v1} Kats and Petrov put forth the first consistent example of a relativistic
quantum field theory which violates the KSS viscosity bound\footnote{See also \cite{v2}.}
--- the $\caln=2$ $Sp(N)$ superconformal gauge theory plasma with 4 fundamental and
1 antisymmetric hypermultiplets. The violation of the viscosity bound can be traced back
to the inequality between the  central charges of the theory, $c\ne a$. More precisely, the bound is
violated once $c-a >0$, which is generic in superconformal gauge theories with $c\ne a$ \cite{v3}.
Moreover, since $c-a \sim N$, this is a finite $N$ correction, and is not due to having finite 't Hooft
coupling.
Once again, the inequality for the central charges on the gauge theory side translates into particular higher-derivative
corrections to the supergravity approximation \cite{bng}, which, to insure reliable computations,
have to be regarded as being 'small'. As a result,  the KSS bound violation
in holographic models realized in string theory is necessarily perturbative.

The work \cite{v1} convincingly established that the original KSS bound
\eqref{bound} can not be a quantitative formulation of a loose quantum-mechanical bound
\eqref{bound0}. Thus, the question remained as to whether or not a
bound of the type \eqref{bound0} existed. As we mentioned above, because of the
universality of the shear viscosity in the supergravity approximation,
any finite violation of the KSS bound  has to be studied in a holographic {\it model}
of the AdS/CFT correspondence, rather than a particular realization of the
holographic correspondence in string theory.
A simple enough model to fulfill this purpose is that of Gauss-Bonnet gravity with a negative
cosmological constant \cite{v2}:
\begin{equation}
S_{GB}=\frac{1}{2l_P^3}\int d^5x \sqrt{-g}\left[ R+ \frac{12}{L^2} +
\frac{\l_{GB}}{2}\ L^2 \left(R^2-4 R_{\mu\nu}R^{\mu\nu}+R_{\mu\nu\r\l}R^{\mu\nu\r\l}\right) \right]\,.
\eqlabel{gbg}
\end{equation}
Up to field redefinitions, for $\l_{GB}\ll 1$ the gravitational model \eqref{gbg}
is equivalent to the string theory holographic example of Kats and Petrov \cite{v1},
for sufficiently large 't Hooft coupling, where one identifies
\begin{equation}
\frac{c-a}{c}=4\l_{GB}+\calo\left(\l_{GB}^2\right) \, .
\eqlabel{gbi}
\end{equation}
The advantage of \eqref{gbg} compared to \cite{v1} is that the former gravitational model
is consistent  for arbitrary values of $\l_{GB}>\frac 14$ \cite{v2}. As such,
it {\it defines} via the AdS/CFT correspondence a dual conformal gauge theory plasma, which we call
 GB plasma,  with central charges \cite{mam}
 \beqa
c&=&\frac{\pi^2}{2^{3/2}}\,\frac{L^3}{\ell_P^3}\,
(1+\sqrt{1-4\lgb})^{3/2}\,\sqrt{1-4\lgb}\,,\nonumber\\
a&=&\frac{\pi^2}{2^{3/2}}\,\frac{L^3}{\ell_P^3}\,
(1+\sqrt{1-4\lgb})^{3/2}\,\left(3\sqrt{1-4\lgb}-2\right)\,,
 \labell{central}
 \eeqa
and hence
 \beq
\frac{c-a}{c}=2\left(\frac{1}{\sqrt{1-4\lgb}}-1\right)\,.
\eqlabel{defd}
 \eeq
Notice the parallel with the construction of \cite{c}: we identified a relativistic
quantum field theory as a holographic dual to \eqref{gbg}, with a shear viscosity
to entropy density ratio given by \cite{v2}
\begin{equation}
\frac{\eta}{s}=\frac{1}{4\pi} \left(1-4\lgb\right) \, ,
\eqlabel{gbv}
\end{equation}
which apparently leads to an arbitrary violation of the KSS bound
(or any bound of the type \eqref{bound0}), given appropriate choices of
$\lgb$ (or equivalently of the central charges of the theory).
To complete the analysis one needs to address the question of the consistency
of the GB plasma, as a relativistic quantum field theory. This was done in
\cite{rob1,bm}. It was found that once
\begin{equation}
\lgb> \frac{9}{100}\,,
\eqlabel{b1}
\end{equation}
the spectrum of excitations in the GB plasma contains modes that propagate
faster than the speed of light \cite{rob1}. Likewise, for
\begin{equation}
\lgb< -\frac{7}{36} \, ,
\eqlabel{b2}
\end{equation}
the GB  plasma also contains microcausality violating excitations \cite{bm}.
Given \eqref{b1} and \eqref{b2} we are led to conclude that consistency
of the GB plasma as a relativistic QFT constrains its viscosity ratio to 
be\footnote{See \cite{Ge:2008ni,Ge:2009eh,Ge:2009ac} for further studies.}
\begin{equation}
\frac{16}{9}\ \ge\ 4\pi \frac{\eta}{s}\ \ge\ \frac{16}{25} \, .
\eqlabel{visco}
\end{equation}
Exactly the same constraint arises by requiring ``positivity of energy''
measured by a detector in the GB plasma \cite{hof}.

To summarize, the example of the GB plasma appears to suggest a link between
the violation of the shear viscosity bound of the type \eqref{bound0} and
the violation of microcausality/positivity of energy in the
theory\footnote{Further work exploring and generalizing this link
appeared in \cite{gc0,gc1,gc2,gc3,gc4}.}.
In this paper we argue that such a link can not be of fundamental nature.

Indeed, the shear viscosity is one of the coupling coefficients of
the effective hydrodynamic description of the theory at lowest momenta
and frequency,  \ie, for
\begin{equation}
{\rm IR:}\qquad\qquad {\w}\ll \min({T,\mu,\cdots})\,,\qquad {|\vec{k}|}\ll \min({T,\mu,\cdots})\,,
\eqlabel{hydro}
\end{equation}
where $\cdots$ stand for any microscopic scales of the plasma, other than temperature
and chemical potential(s) for the conserved charge(s).
On the contrary, the microcausality of the theory is determined by the propagation
of the modes in exactly the opposite regime, \ie, for
\begin{equation}
{\rm UV:}\qquad\qquad {\w}\gg \max({T,\mu,\cdots})\,,\qquad {|\vec{k}|}\gg \max({T,\mu,\cdots}) \, .
\eqlabel{causality}
\end{equation}
A link between the features of the theory governing its microcausality
and its shear viscosity is only possible if the {\it same phase} of the theory
extends over the entire range of the energy scales --- from the infrared to the ultraviolet.
In other words, there must not be any phase transitions in the plasma.
This is precisely what is happening in the GB plasma! Since the GB plasma is conformal,
and temperature is the only available scale in the model, there can not be
any phase transition in the theory as a function of temperature.
The only free parameter of the model is the GB coupling constant $\lgb$,
which determines both the shear viscosity ratio and its microcausality
properties. Hence the link between the two, originally found in \cite{rob1},
is not surprising --- rather, in a sense, it is an accident.

Consider a conformal plasma in the presence of chemical potential, defined
as a holographic dual to appropriately --- see the next section for details
--- generalized GB
gravity. Assume that this plasma undergoes a second order phase transition
below some critical temperature $T_c\propto \mu$ associated with the spontaneous
breaking of some global  $U(1)$ symmetry and the generation of a condensate of some
irrelevant operator $\calo_c$:
\begin{equation}
\langle \calo_c\rangle \
\begin{cases}
=0\,,\qquad T>T_c\\
\ne 0\,,\qquad  T<T_c\,.
\end{cases}
\eqlabel{cond}
\end{equation}
Clearly, if the model is engineered in such a way that the effective GB coupling
of the dual gravitational description is
\begin{equation}
\lgb\bigg|^{effective}\ \propto\  \calo_c \, ,
\eqlabel{leff}
\end{equation}
it is natural to expect given \eqref{cond} that
\begin{equation}
\lgb\bigg|^{effective}\
\begin{cases}
=0\,,\qquad {\rm UV}\\
\ne 0\,,\qquad  {\rm IR }\,.
\end{cases}
\eqlabel{luvir}
\end{equation}
In such a model, microcausality features -- governed by the unbroken phase --
would be completely decoupled from the physics
that determines the shear viscosity of the symmetry-broken phase of the
plasma.
Also, as a result, the shear viscosity to entropy density ratio in the UV would
differ from that in the IR.
Thus, although $\eta/s$ does not flow in any Wilsonian sense (see e.g. \cite{Bredberg:2010ky}),
in this construction the decoupling of the UV physics from the IR is reflected
in the different behavior of $\eta/s$ in the two regimes.

In the next section we present a detailed holographic model of AdS/CFT correspondence
implementing the ``decoupling idea'' outlined above.
We study the thermodynamics of the model in section 3. The results of
the ratio of the shear viscosity to the entropy density
and the causality analysis of the model
are discussed in sections 4 and 5, respectively. We conclude in section 6.

\section{The holographic model}
Following the general idea presented in the introduction, we would like to engineer
a holographic model of AdS/CFT with a spontaneous symmetry breaking in the IR
and non-universal shear viscosity in the symmetry-broken phase.

Our starting point is the holographic model of superfluidity proposed in \cite{gub1}\\ (GHPT)
and described by
\begin{equation}
\call_{superfluid}=R-\frac{L^2}{3} F_{\mu\nu}F^{\mu\nu}+\left(\frac{2L}{3}\right)^3\frac 14
\epsilon^{\lambda\mu\nu\sigma\rho}F_{\lambda\mu}F_{\nu\sigma}A_\rho
+\call_{scalar}^{SUGRA} \, ,
\eqlabel{ssuper}
\end{equation}
with
\begin{equation}
\call_{scalar}^{SUGRA}=-\frac{1}{2}\left[(\del_\mu\phi)^2+\sinh^2\phi\ (\del_\mu\theta-
2A_\mu)^2-\frac{6}{L^2}\ \cosh^2\frac\phi 2\ (5-\cosh\phi)\right] \, .
\eqlabel{lscalar}
\end{equation}
Here, $\phi$ and $\theta$ are the modulus and the phase of a complex scalar $\Psi$,
which is dual to a chiral primary operator $\calo$ with scaling dimension $\Delta$.
By a $U(1)$ gauge transformation we can set $\theta=0$.
The model \eqref{ssuper} is a consistent truncation of type IIB
supergravity, and represents a string holographic realization of the
mean-field second-order phase transition.

We now briefly summarize the features and the dynamics of the model \eqref{ssuper}\footnote{
See  \cite{gub1} for further details.}.
The gauge field $A_\mu$ is dual to a global $U(1)$ R-symmetry current,
and has been normalized in such a way
that chiral primaries have R-charge $|R|=2\Delta/3$ \cite{Berenstein:2002ke}.
By expanding the scalar potential to quadratic order in $\phi$,
\beq
V(\phi) = -\frac{12}{L^2} - \frac{3}{2L^2} \phi^2 + \ldots\,,
\eeq
we can read off the mass of the scalar $m^2 L^2 = \Delta (\Delta-4)=-3$,
and extract the dimension of the dual operator.
Thus, we can identify $\Psi$ with a chiral primary operator $\calo$ of dimension $\Delta=3$
and R-charge $R=2$. Since the non-normalizable component of $\phi$
is set to zero, the dual QFT is a conformal gauge theory.

Consider this gauge theory at finite temperature $T$ and nonzero chemical potential $\mu$.
It was found in \cite{gub1} that for $T<T_c\approx 0.0607\mu$  the GHPT plasma undergoes a mean-field
second-order phase transition associated with the development of the condensate for $\calo$:
\begin{equation}
\langle\calo\rangle\ \begin{cases}
=0\,,\qquad T>T_c\\
\propto (T-T_c)^{1/2}\,,\qquad T\le T_c \, .
\end{cases}
\eqlabel{ocond}
\end{equation}
On the gravity side, while at high temperatures the background is that of an electrically charged $AdS$ black hole,
once the temperature drops below $T_c$ the black hole develops scalar hair.
Since $\calo$ is charged under the global $U(1)$ symmetry, the
condensation breaks this symmetry spontaneously.
While the precise value of the critical temperature is sensitive to the details
of the full scalar Lagrangian $\call_{scalar}^{SUGRA}$ (dual to $\calo$) in the gravitational description,
the existence of the  transition itself depends only on the set of values
$(R,\Delta)$ of the operator in question \cite{har}. Thus, for the purpose of
engineering the phase transition only --- we are going to give up the string theory embedding
anyway --- we simplify
\begin{equation}
\call_{scalar}^{SUGRA}\to \call_{scalar}=-\frac 12\left[\left(\del_\mu\phi\right)^2+4 \phi^2 A_\mu A^\mu\right]+\frac{12}{L^2}+\frac{3}{2L^2}\phi^2\,,
\eqlabel{lscalar1}
\end{equation}
while maintaining $(R=2,\Delta=3)$ for the dual operator $\calo$.

So far, while the simplified $\call_{superfluid}$ describes a second-order phase transition, the universality
theorem of \cite{bu} guarantees that the shear viscosity to entropy density ratios of the low-temperature
(symmetry broken) and the high-temperature (symmetry unbroken)  phases in this plasma are the same as in \eqref{univ}.
Furthermore, since in the UV the asymptotic geometry described by $\call_{superfluid}$ is the same as that of the
Reissner-Nordstrom black hole in $AdS_5$ \cite{gub1}, the causality properties of the dual plasma must be identical to
those of $\caln=4$ SYM plasma. In particular, we do not expect any violation of
microcausality\footnote{We explicitly verify this in section 5.}.

To proceed, we need to introduce higher-derivative gravitational corrections into $\call_{superfluid}$ in such a way that:
\nxt the resulting equations of motion for the background and the fluctuations are always of second order;
\nxt these corrections must vanish in the symmetric phase, while being nonzero in the symmetry-broken phase;
\nxt the phase transition itself should not be destroyed by these corrections.

A natural modification, obviously satisfying the constraints above, is achieved by generalizing the Gauss-Bonnet
coupling $\lgb$ of the higher-derivative term in \eqref{gbg} as
\begin{equation}
\frac\lgb 2 \ \to\ \b\ (\Psi \Psi^*)^n=\b\ \phi^{2n}\,,\qquad n\ge 2\,,
\eqlabel{lgbmod}
\end{equation}
for some fixed coupling constant $\b$. Indeed, the Gauss-Bonnet combination leads to second-order equations of motion.
In the symmetric phase the expectation value of $\calo_c\equiv (\Psi \Psi^*)^n$ vanishes, suggesting that the UV properties of the theory
must be exactly as for $\b=0$; the $n\ge 2$ condition guarantees that the mass of $\Psi$ (and thus the dimension
of the operator $\calo$) will not change as $\b\ne 0$. Finally, the sign of $\b$ will control whether
the shear viscosity ratio in the symmetry-broken phase is above or below the universal result \eqref{univ}.

We can now present our model:
\begin{equation}
\call=R-\frac{L^2}{3} F_{\mu\nu}F^{\mu\nu}+\left(\frac{2L}{3}\right)^3\frac 14 \epsilon^{\lambda\mu\nu\sigma\rho}F_{\lambda\mu}F_{\nu\sigma}A_\rho
+\call_{scalar}+\call_{GB} \, ,
\eqlabel{ea1}
\end{equation}
where $\call_{scalar}$ is given by \eqref{lscalar1}, and
\begin{equation}
\call_{GB}=\b \phi^4 L^2\biggl(R^2-4 R_{\mu\nu}R^{\mu\nu}+R_{\mu\nu\r\l}R^{\mu\nu\r\l}\biggr) \, .
\end{equation}
Thus, while in the UV the scalar field is turned off and one has the simple Einstein-Maxwell two-derivative theory,
at low energies the scalar field condenses, and controls the strength of the higher-derivative GB correction.
Note that we have set $n=2$ in (\ref{lgbmod}).
Finally, we will consider the dynamics of \eqref{ea1} while taking the non-normalizable component of $\phi$ to be zero. Thus,
$\call$ defines holographically a dual \emph{conformal} gauge theory plasma with a global $U(1)$ symmetry.

To describe the equilibrium state of the plasma -- dual to \eqref{ea1} -- at finite
temperature and in the presence of a $U(1)$ chemical potential  we take the following ansatz for the background fields:
\begin{equation}
\begin{split}
&ds_5^2=-c_1^2 dt^2+c_2^2 d\vec{x}^2+c_3^2 dr^2\,,\quad A_\mu=A \, \delta^0_\mu\,,\\
&c_1=\frac{z_0\sqrt{f}}{\sqrt{r}}\,, \quad c_2=\frac{z_0}{\sqrt{r}}\,,\quad c_3=\frac{g}{2\sqrt{f}r}\,,\quad A=\a z_0 \, .
\end{split}
\eqlabel{symfields}
\end{equation}
where $\{f\,, g\,, \a\,, \phi\}$ are functions of the radial coordinate $r$ only.
Without loss of generality we can choose this radial coordinate such that
 $r=0$ corresponds to the boundary while $r=1$ is the location of the horizon, \ie
\begin{equation}
\lim_{r\to 1_-} c_1=0\,,\qquad \lim_{r\to 1_-} c_2={\rm finite}\equiv z_0\,,\qquad \lim_{r\to 0_+}\frac {c_1}{c_2}=1\,,
\eqlabel{boho}
\end{equation}
for an arbitrary constant $z_0$. Lastly, we set $L=1$.

\section{The background geometry}

In this section we discuss the thermodynamics of the holographic model
\eqref{ea1}-\eqref{symfields}.
It is straightforward to derive the equations of motion for
the background fields $\{f,g,\a,\phi\}$ --- in the parametrization
\eqref{symfields}, we find two second
order equations for $\{\a,\phi\}$, and two first order equations for
$\{f,g\}$\footnote{These equations are too long to be presented here.
They are available from the authors upon request.}.

The asymptotic solution near the boundary is given by
\begin{equation}
\begin{split}
\a =& \a_0+\a_1\ r+\frac 14 p_1^2\a_0\ r^3+\calo(r^4) \, ,\\
\pp =& p_1\ r^{1/2}\ \left(r-\frac 12 \a_0^2 r^2+\left(\frac{1}{12}\a_0^4
-\frac 13 \a_0\a_1-\frac 38 f_2\right)\ r^3+\calo(r^4)\right) \, , \\
f =& 1+f_2\ r^2+\frac 29 \a_1^2\ r^3+\calo(r^4)\, , \\
g =& 1-\frac 14 p_1^2\ r^3+\calo(r^4) \, .
\end{split}
\eqlabel{uv}
\end{equation}
Thus, at the boundary the metric reduces to the simple form:
\beq
ds^2=\frac{z_0^2}{r}(-dt^2+d\vec{x}^2) + \frac{dr^2}{4 r^2} \, .
\eeq
Note that we have set the non-normalizable component of $\phi$ to zero,
since we are discussing spontaneous symmetry breaking in conformal gauge theories.
Altogether the UV asymptotics are determined by 4 parameters:
$\{\a_0,\a_1,p_1,f_2\}$. Of these, the first one, namely $\a_0$,
is the coefficient of the non-normalizable mode related to the $U(1)_R$ chemical potential $\mu$, while the rest
are related to the expectation values of various operators. The parameter
%(see below for details).
$\a_1$, for instance, is the charge density conjugate to the chemical potential.

The asymptotic solution near the horizon $y\equiv 1-r$ is given by
\begin{equation}
\begin{split}
\a=&a^h_1\ y+\calo(y^2) \, ,\\
\pp=&p^h_0+\calo(y)\, ,\\
f=&\calo(y)\, ,\\
g=&g^h_0+\calo(y) \, ,
\end{split}
\eqlabel{ir}
\end{equation}
where we indicated only the independent parameters. Thus, in the IR altogether we have
3 independent parameters: $\{a^h_1,p^h_0,g^h_0\}$.

The temperature $T$ and chemical potential $\mu$ are
\beqa
\label{temp}
T &=& \frac{z_0(72 (g^h_0)^2+9 (g^h_0 p^h_0)^2-8 (a^h_1)^2)}{72\pi g^h_0}\,,\\
\mu &=& z_0 \, \a_0 \,.
\eeqa
The thermodynamic potentials are given by\footnote{These expressions
can be obtained following the same procedure as in \cite{bl}.}
\begin{equation}
\begin{split}
\om=&-P=\frac{1}{2l_P^3}\ \left(z_0^4 f_2\right)\,, \qquad \cale=3 P \, ,\\
s T=&\frac{1}{2l_P^3}\ \frac{4z_0^4(\a_0\a_1-3f_2)}{3}\,,\qquad \r=-\frac{1}{2l_P^3}\
\frac 43 z_0^3 \a_1 \, ,
\end{split}
\eqlabel{thermo}
\end{equation}
with $\om$ denoting the Gibbs free energy.
When translating to gauge theory variables, we identify
\begin{equation}
\frac{1}{2l_P^3}=\frac{N^2}{8\pi^2} \, ,
\eqlabel{defg5}
\end{equation}
as in the case of $\caln=4$ SYM.

Note that the expression for the entropy density in \eqref{thermo} was derived
imposing the basic thermodynamic relation
\begin{equation}
\om=\cale -s\ T -\mu\ \r \, .
\eqlabel{basic}
\end{equation}
Alternatively, the entropy density can be computed using  Wald's entropy formula \cite{Wald:1993nt},
\beq
S = - 2 \pi \int_{\Sigma} d^{D-2} x \sqrt{- h} \, \frac{\delta {\mathcal L}}{\delta R_{\mu \nu \rho \sigma}}
\, \epsilon_{\mu \nu} \epsilon_{\rho \sigma} \, ,
\label{wald}
\eeq
where $\Sigma$ denotes the horizon cross section, $h$ is the induced metric on it and $\epsilon_{\mu \nu}$ is the
binormal to the horizon cross section.
For our geometry the binormal is $\epsilon_{tr}=c_1 c_3$, obeying $\epsilon_{\mu\nu} \epsilon^{\mu\nu}=-2$, and
\beq
2 l_P^3\ \frac{\delta \mathcal{L}}{\delta R_{\mu \nu \rho \sigma}} =
 g^{\mu\rho}g^{\nu\sigma} + 2 \beta \phi^4 L^2 \left( R \, g^{\mu \rho}g^{\nu \sigma} -4 g^{\mu\rho} R^{\nu\sigma} + R^{\mu\nu\rho\sigma} \right) \, .
\eeq
Putting all the various ingredients together, we find that
\beq
S= -2\pi A_h \frac{\delta \mathcal{L}}{\delta R_{\mu\nu\rho\sigma}}  \, \epsilon_{\mu\nu} \epsilon_{\rho \sigma} =
\frac{1}{2l_P^3}\ A_h \left. \left( 4\pi - 48 \pi \beta\phi^4 L^2 \frac{(\partial_r c_2)^2}{c_2^2 c_3^2} \right) \right|_{horizon}\,,
\eqlabel{sw}
\eeq
where $A_h$ denotes $\int_{\Sigma} d^{D-2} x \sqrt{- h}$.
Given \eqref{symfields}, from \eqref{sw} we find that the entropy density is
\begin{equation}
s\big|_{Wald}=\frac{1}{2l_P^3}\ 4\pi \, z_0^3 \, .
%  s\big|_{Wald}=\frac{1}{2l_P^3}\ z_0^3 \, .
\eqlabel{swald}
\end{equation}
We mention in passing that a highly nontrivial consistency check on our numerical data would be the
agreement of the entropy density in \eqref{thermo} with the one in \eqref{swald}. We will return to this point later
in this section.

\subsection{Symmetric phase}

In the symmetric phase the field $\pp$ is identically zero, which tells us that the parameters $p_1$ and $p_0^h$ vanish.
Thus, once  $\{T,\mu\}$ are fixed --- \emph{i.e.} given $\{z_0,\a_0\}$ --- we are left
with 4 integration constants $\{\a_1,f_2,a_1^h,g^h_0\}$ ---
precisely the correct number necessary to uniquely solve a coupled system of
1 second-order differential equation (for $\a$) and 2 first-order
differential equations (for $\{f,g\}$). Actually, in this case the background equations of motion can be solved
analytically\footnote{Our numerical results are in excellent agreement with the exact
analytical result.}.
We find:
\begin{equation}
\begin{split}
\a=\a_0(1-r)\,,\qquad f=1-r^2+\frac{2 \a_0^2}{9}(r^3-r^2)\,,\qquad g=1 \, .
\end{split}
\eqlabel{syssol}
\end{equation}

In this case the thermodynamics is that of the $\caln=4$ SYM plasma with
the same chemical potentials for all the  $U(1)^3\subset SU(4)$ R-symmetry
global charges \cite{Behrndt:1998jd}.
The expression for the temperature reduces to
\beq
T=\frac{z_0}{\pi} \left( 1-\frac{\alpha_0^2}{9}\right)\, ,
\eqlabel{tsym}
\eeq
in terms of which the entropy density becomes
\beq
s=\frac{1}{2 l_P^3} \frac{4 \pi^4 \, T^3}{\left(1-\frac{\alpha_0^2}{9}\right)^3}  \, .
\eeq

\subsection{Broken phase}

\begin{figure}[t]
\begin{center}
\psfrag{tred}{{$\frac{2\pi T}{\mu}$}}
\psfrag{fred}{{$\frac{8}{N^2\pi^2}\ \frac{\om}{T^4}$}}
\psfrag{fr}{{${\om_{blue}}/{\om_{symmetric}}$}}
  \includegraphics[width=3in]{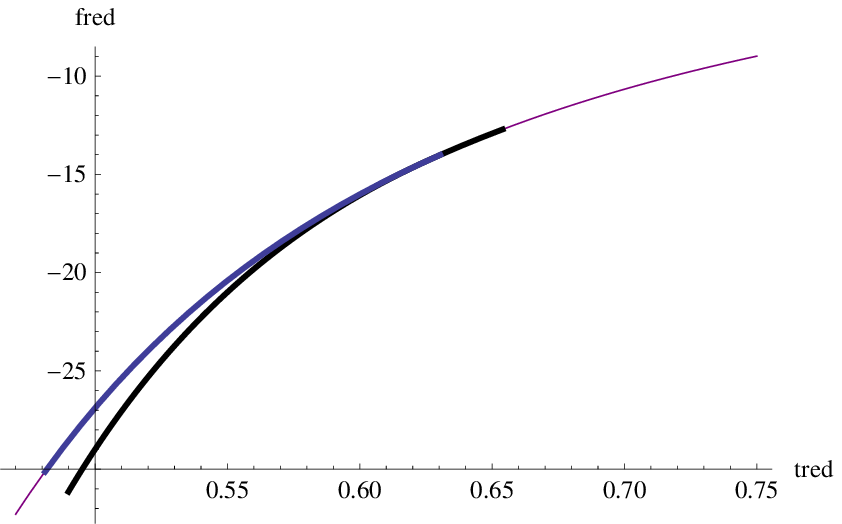}
  \includegraphics[width=3in]{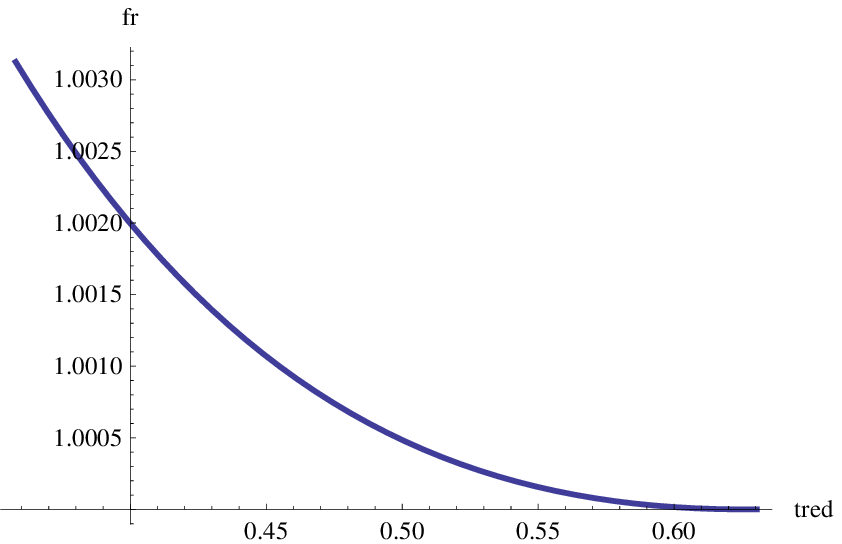}
\end{center}
  \caption{(Colour online) Left plot: the Gibbs free energy densities
of the symmetric phase (purple curve) and the symmetry-broken
phases at $\b=0$ (black curve) and $\b=-1$ (blue curve)
as a function of $\frac{2\pi T}{\mu}$.
Right plot: the ratio of the Gibbs free energies in the symmetric and
the broken phases at $\b=-1$.
}
%\caption{(Colour online) The Gibbs free energy densities
%of the symmetric phase (purple curve) and the symmetry-broken
%phases at $\b=0$ (black curve) and $\b=-1$ (blue curve)
%as a function of $\frac{2\pi T}{\mu}$ --- left plot.
%The ratio of the Gibbs free energies in the symmetric and
%the broken phases at $\b=-1$ --- right plot.}
\label{figure1}
\end{figure}

\begin{figure}[t]
\begin{center}
\psfrag{tred}{{$\frac{2\pi T}{\mu}$}}
\psfrag{fred}{{$\frac{8}{N^2\pi^2}\ \frac{\om}{T^4}$}}
\psfrag{fbfs}{{${\om_{broken}}/{\om_{symmetric}}$}}
  \includegraphics[width=4.5in]{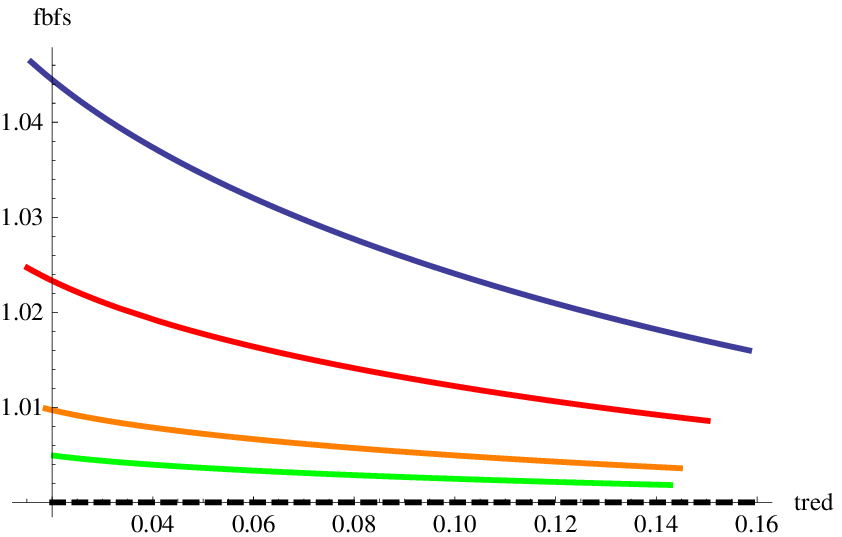}
\end{center}
  \caption{(Colour online) Ratios of the Gibbs free energy densities in the
broken and unbroken phases as a function of $\frac{2\pi T}{\mu}$, for select values of the coupling.
From top to bottom: $\b=-1$ (blue), $\b=-2$ (red), $\b=-5$ (orange) and $\b=-10$ (green).
} \label{figure2}
\end{figure}

In the broken phase the field $\pp$ is no longer zero, and the parameters $p_1,p_0^h$ are now turned on.
Thus, for a fixed  $\{T,\mu\}$ --- given $\{z_0,\a_0\}$ --- we are left
with 6 integration constants $\{\a_1,p_1,f_2,a_1^h,p_0^h,g^h_0\}$. This is
precisely the correct number necessary to uniquely solve a coupled system of
2 second-order differential equations (for $\{\a,\pp\}$) and 2 first-order
differential equations (for $\{f,g\}$). We use numerical techniques developed
in \cite{abk} to study the thermodynamics of the low-temperature symmetry-broken
phase of \eqref{ea1} for different values of the coupling constant $\b$.

We find that the mean field second-order phase transition at $\b=0$
persists for $\b\ne 0$. The positive values of $\b$ tend to increase the ratio
$\frac{T}{\mu}$ (for fixed non-normalizable modes $\{z_0,a_0\}$), while the negative
values of $\b$ tend to decrease it. We have not performed an exhaustive analysis of the
phase diagram of the system, but rather identified interesting values of the coupling with regards to
the ratio of shear viscosity to entropy density (see section 4).

A  representative case of this analysis is the comparison
between the Gibbs free energy densities of the broken and
unbroken phases as a function of $\frac{2\pi T}{\mu}$ at $\b=\{0,-1\}$, which is shown in
Fig.~\ref{figure1}. On the left, the ``thin purple'' curve represents the Gibbs free
energy density of the symmetric phase
\begin{equation}
\frac{8}{\pi^2 N^2}\ \frac{\om}{T^4}\equiv f_{purple}\left(x\equiv \frac{2\pi T}{\mu}\right)
=-\frac{1024}{27}\ \frac{3x^2+4-x\sqrt{9x^2+16}}{x^4(\sqrt{9x^2+16}-3x)^4} \, .
\end{equation}
On the other hand the ``thick black''and the "thick blue'' curves represent the Gibbs free energy
densities of the broken phase at, respectively, $\b=0$ and $\b=-1$.
As the temperature increases, the condensate of the dimension-3 operator
$\langle \calo_3\rangle$
dual to the holographic scalar $\phi$ in \eqref{lscalar1}
\begin{equation}
\langle \calo_3\rangle\propto p_1
\eqlabel{vev3}
\end{equation}
decreases, ultimately vanishing at some critical temperature $T_c=T_c(\b)$.
We find
\begin{equation}
\frac{2\pi T_c}{\mu}\bigg|_{\b=0}=0.65396(3)\,,\qquad
\frac{2\pi T_c}{\mu}\bigg|_{\b=-1}=0.63040(9) \, .
\eqlabel{crit}
\end{equation}
The right plot in Fig.~\ref{figure1} represents the ratio of the
free energies in the broken and unbroken phases at $\b=-1$.
Clearly, the broken phases are thermodynamically favorable at low
temperatures.

To get to the interesting regime in the shear viscosity ratios  $\frac{\eta}{s}$
(in the broken phases)
we need to get to temperatures several times smaller than the appropriate
critical temperature. Fig.~\ref{figure2} shows ratios of the
Gibbs free energy densities in the broken and unbroken phases for a select
set of couplings, $\b=\{-1,-2,-5,-10\}$.  The broken phases, while being closer and closer
to the unbroken phase as $\b$ decreases,  are thermodynamically
preferable for each given temperature.

An important consistency check on our numerical analysis is the
comparison\footnote{Of course, this is a nontrivial check in the symmetry
broken phases only.} of the
entropy density derived from the basic thermodynamic relation  \eqref{basic} ---
see \eqref{thermo} --- and the one obtained directly from the Wald's entropy
--- see \eqref{swald}. In all instances we find
\begin{equation}
\bigg|\ \frac{s}{s|_{Wald}}-1\ \bigg|< 10^{-6} \, .
\eqlabel{sratio}
\end{equation}

\section{Shear Viscosity}

In the hydrodynamic approximation, linear response theory implies that the retarded Green's function
of the stress energy tensor of the conformal fluid in the
%the (hydrodynamic approximation in the)
tensor channel is given by \cite{andrei2}
\begin{equation}
\begin{split}
&G_{R}^{xy,xy}\left(\ww\equiv \frac{\w}{2\pi T},\vec k=0\right)
=-i \int dt \,d\vec{x} \, e^{i\omega t} \theta(t) \,
\langle [T_{xy}(x),T_{xy}(0)] \rangle\\
&=P\left(1-2\pi\ww i\ \frac{\eta}{s}\ \frac{P+\cale-\mu\r}{P}+\calo(\ww^2)\right) \, .
\end{split}
\eqlabel{gr}
\end{equation}
Techniques for computing this correlation function in a dual gravitational model
are well developed \cite{ss}, and we will not review them here. In particular, the analysis which we
perform in this note is equivalent to that in \cite{bls}.

On the gravity side, computation of the Green's function
\eqref{gr} entails adding a metric perturbation of the form
\beq
g_{xy} \rightarrow g_{xy} +  h_{xy} \, ,
%\quad h_{xy} = \epsilon \, \varphi(t,r,z)
\eeq
and finding the effective action for the fluctuation $h_{xy}$.
Thus, we take the metric to be
\beq
ds_5^2=-c_1^2 dt^2+c_2^2 (dx^2+dy^2+dz^2+2 \, \epsilon \, \Phi \, dx dy)+c_3^2 dr^2\,,
\eeq
and expand
\beq
\Phi(t,r,z) = \int \frac{d^4 k}{(2\pi)^4} \, e^{-i\omega t + i k z} \varphi_k(r) \, .
\eeq
Since we are ultimately interested in the correlator at vanishing spatial momentum, we can
set $k=0$ at this stage and consider perturbations which depend on $(r,t)$ only.
Expanding the action (\ref{ea1}) to second order in the perturbations,
%(i.e. second order in $\epsilon$),
we can easily see
that the effective action for the fluctuation is of the form originally found in
\cite{bls},
\beqa
S_\varphi^{(2)} &=& \frac{1}{2 l_P^3} \int \frac{d^4 k}{(2\pi)^4} \, dr
\Bigl[A(r) \varphi_k^{\prime\prime}\varphi_{-k} + B(r) \varphi_k^{\prime}\varphi_{-k}^\prime + C(r) \varphi_k^{\prime}\varphi_{-k}
+D(r) \varphi_k \varphi_{-k} + \nonumber \\
&& \quad \quad \quad \quad +  E(r) \varphi_k^{\prime\prime} \varphi_{-k}^{\prime\prime} + F(r) \varphi_k^{\prime\prime}
\varphi_{-k}^{\prime}\Bigr] + \mathcal{K}_{GB}+\calk_{counter} \, .
\eeqa
Here $\mathcal{K}_{GB}$ denotes the generalized Gibbons-Hawking boundary term, needed to ensure a
well-defined variational principle, and $\calk_{counter}$ is a local boundary counterterm,
necessary to remove UV divergences in the stress energy tensor correlation functions\footnote{For the
construction of the generalized Gibbons-Hawking boundary terms and counterterms for
Einstein-Maxwell theory in the presence of generic $R^2$ corrections, see \cite{Cremonini:2009ih}.}.

For our model (\ref{ea1}) we find that $E(r)=0$, which is in agreement with the expectation
that the equation of motion corresponding to Gauss-Bonnet gravity should not contain more
than two derivatives.
When $E=0$ the generalized Gibbons-Hawking term takes the simple form
\beq
\mathcal{K}_{GB} = \int_{\partial M} \frac{d^4 k}{(2\pi)^4}
\Bigl[ -A \varphi_k^{\prime}\varphi_{-k} - \frac{F}{2} \varphi_k^{\prime}\varphi_{-k}^\prime \Bigr] \, .
\eeq
Furthermore, the local boundary counterm is precisely as in the case of pure $AdS_5$ \cite{bk}:
\beq
\calk_{counter}=\int_{\partial M} \frac{d^4 k}{(2\pi)^4}
\Bigl[\ \frac{\calb}{2} \varphi_k\varphi_{-k}\ \Bigr] \, .
\eeq

It turns out to be particularly convenient to rewrite the action in the form
\beqa
S_\varphi^{(2)} &=& \frac{1}{2 l_P^3} \int_M \frac{d^4 k}{(2\pi)^4} \, dr
\Bigl[\Bigl(B-A-\frac{F^\prime}{2}\Bigr) \varphi_k^{\prime}\varphi_{-k}^\prime +
\Bigl(D-\frac{C^\prime-A^{\prime\prime}}{2}\Bigr)
\varphi_k \varphi_{-k} \Bigr] + \nonumber \\
&+& \frac{1}{2 l_P^3} \int_{\partial M} \frac{d^4 k}{(2\pi)^4} \half (C-A^\prime+\calb) \varphi_k \varphi_{-k} \, ,
\eeqa
from which one can easily read off the \emph{radial} canonical momentum for the scalar $\varphi$,
\beq
\Pi_k(r) \equiv \frac{\delta S^{(2)}_\varphi}{\delta \varphi_{-k}^\prime} =
\frac{1}{l_P^3} \Bigl(B-A-\frac{F^\prime}{2}\Bigr) \varphi_{k}^{\, \prime}  \, .
\eeq
Introducing an ``effective mass'' term for the scalar fluctuation
\beq
M_{eff}(r) \equiv \frac{1}{l_P^3} \Bigl(D-\frac{C^\prime-A^{\prime\prime}}{2}\Bigr),
\eeq
the scalar equation of motion can be written in the simple form
\beq
\partial_r \Pi_k = M_{eff}  \varphi_k \, .
\eqlabel{eoms}
\eeq
By making use of the background equations of motion it is straightforward to verify that
$M_{eff}=\calo(\ww^2)$,  which in turn means that
the radial flow of $\Pi$ is \emph{trivial} in the $\ww \rightarrow 0$ limit, and
the mass term $M_{eff}$ does not contribute to \eqref{gr} to order $\calo(\ww)$.

Finally, we note that evaluating the on-shell action to order $\calo(\ww)$
turns out to be equivalent to evaluating the following boundary
term\footnote{Explicit expressions for $\{A,B,C,F,\calb\}$ are too long to be presented here.
They are available from the authors upon request.}
\beqa
&&
S_{\text{on-shell}} =  \int \frac{d^4 k}{(2\pi)^4} \, \mathcal{F}_k \nonumber \\
&=& \frac{1}{2 l_P^3} \left. \int \frac{d^4 k}{(2\pi)^4}\left[ \left(B-A-\half F^\prime  \right)\varphi_{k}^\prime \varphi_{-k} +
\half(C-A^{\prime}+\calb) \varphi_{k}\varphi_{-k} \right] \right|_{r=0}^{r=1}\,,
\eeqa
with the \emph{flux} $\mathcal{F}_k$ directly related to the retarded Green's function:
\beq
G^R_{xy,xy} = - \lim_{r \rightarrow 0} \frac{2\mathcal{F}_k}{\varphi_k(r)\varphi_{-k}(r)} \, .
\eqlabel{drf}
\eeq

Much like the background in the broken phase (see section 3), the equation of motion for the scalar $\p_k$ \eqref{eoms}
has to be solved numerically. The fluctuation $\p_k$ must satisfy an incoming wave boundary condition at the horizon
\cite{ain}
\begin{equation}
\p_k=(1-r)^{-i\ww/2}\ \psi_k(r)\,,\qquad \lim_{r\to 1} \psi_k=1 \, ,
\eqlabel{inc}
\end{equation}
where $\psi_k(r)$ is regular near the horizon, $r\to 1_-$.
Note that we used a conventional normalization for $\psi_k$.
To compute the correlator \eqref{drf} to order $\calo(\ww)$, we need to solve \eqref{eoms}
to order $\calo(\ww)$ as well. We represent
\begin{equation}
\psi_k(r)=\psi_k^0(r)+i\ww\ \psi_k^1(r)+\calo(\ww^2)\, .
\eqlabel{p1}
\end{equation}
Demanding regularity at the horizon we find
\begin{equation}
\psi_k^0=1
\end{equation}
identically. The second order linear inhomogeneous equation for $\psi_k^1$
has the following asymptotic solution
\begin{equation}
\begin{split}
\psi_k^1=&\psi_0-\frac 12 r +\psi_{2}\ r^2-\frac 16 r^3-\left(\frac 18 +\frac 18\ f_2+\frac 12\ f_2 \psi_2\right)r^4+\calo(r^5) \, ,\\
\psi_k^1=&\calo(y) \, .
\end{split}
\eqlabel{psi1}
\end{equation}
close to  the boundary $r\to 0$ and the horizon $y\to 0$ correspondingly.
It is uniquely specified by two parameters $\{\psi_0, \psi_2\}$.
Comparing the holographic expression for the Green's function \eqref{drf} with that of
the hydrodynamics \eqref{gr}, we arrive at a fairly simple expression for the
ratio of shear viscosity to entropy density:
\begin{equation}
\frac{\eta}{s}=\frac{3}{8\pi}\ \frac{1+4\psi_{2}}{\a_0 \a_1-3 f_2} \, .
\eqlabel{etas}
\end{equation}

\subsection{Shear viscosity of the symmetric phase}

In the symmetric phase the background is known analytically, and is given by \eqref{syssol}.
The equation of motion for $\psi_k^1$ takes the following form
\begin{equation}
0=\psi_k^{1\,\prime\prime}+\frac{4\a_0^2(r^3-r^2)-9(1+r^2)}{(r^2-r)(2\a_0^2 r^2-9(r+1))}\ \psi_k^{1 \, \prime}-\frac{2\a_0^2 r^2+9}{2(r^2-r)
(2\a_0^2r^2-9(1+r))} \; ,
\eqlabel{eomspsi1}
\end{equation}
which can be solved analytically:
\begin{equation}
\psi_k^1=\int_r^1\ dx\ \frac{2\a_0^2 x+9}{2(9(1+x)-2\a_0^2 x^2)} \, .
\eqlabel{spi1k}
\end{equation}
From \eqref{spi1k} we can extract
\begin{equation}
\psi_2=\frac 14 -\frac{1}{18}\ \a_0^2 \, .
\eqlabel{respsi1}
\end{equation}
Finally, using the explicit solution \eqref{syssol}, relation \eqref{etas}
reproduces the shear viscosity of the $\caln=4$ SYM plasma in the presence of $U(1)_R$ chemical potential \cite{ssc} :
\begin{equation}
\frac{\eta}{s}=\frac{1}{4\pi} \, .
\eqlabel{eqaution}
\end{equation}

\subsection{Shear viscosity of the broken phase }

\begin{figure}[t]
\begin{center}
\psfrag{Tred}{{$\frac{2\pi T}{\mu}$}}
\psfrag{rat}{{$4\pi\ \frac{\eta}{s}$}}
%\psfrag{fbfs}{{${\om_{broken}}/{\om_{symmetric}}$}}
  \includegraphics[width=4.5in]{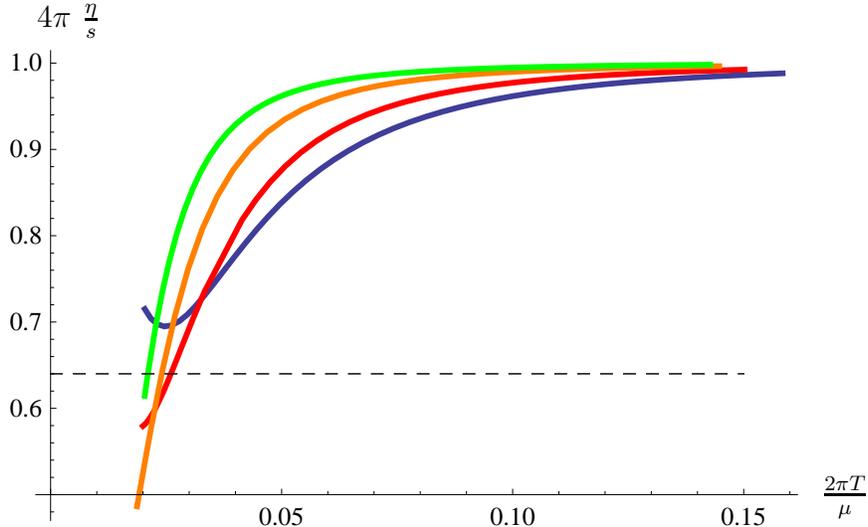}
\end{center}
  \caption{(Colour online) Ratios of shear viscosity to entropy density
in the broken phase for select values of the coupling:
$\b=-1$ (blue), $\b=-2$ (red), $\b=-5$ (orange) and $\b=-10$ (green),
as a function of $\frac{2\pi T}{\mu}$. The dashed black line indicates the Gauss-Bonnet
viscosity bound: $\eta/s\ge 16/25$ \cite{rob1}.
} \label{figure3}
\end{figure}

In the broken phase the equation for motion for $\psi_k^1$ \eqref{p1} and the
shear viscosity to entropy density ratio \eqref{etas} must be computed numerically.
First, for $\b=0$ we find
\begin{equation}
\bigg|4\pi\ \frac{\eta}{s}-1\bigg|_{\b=0}\ <\ 10^{-7} \, ,
\end{equation}
and therefore recover the universal $\eta/s=1/4\pi$ result \cite{bu}  expected
for a two-derivative theory.

Fig.~\ref{figure3} represents the results of the numerical analysis of the shear viscosity
in the symmetry-broken phase of the holographic model \eqref{ea1}, for select values
of the coupling. We show $\b=-1$ (blue), $\b=-2$ (red), $\b=-5$ (orange) and $\b=-10$ (green),
as a function of $\frac{2\pi T}{\mu}$. Notice that for $\b=-1$ the shear viscosity
remains above the causality bound for Gauss-Bonnet gravity found in \cite{rob1}
\begin{equation}
4\pi \, \frac{\eta}{s}\ge \frac{16}{25}=0.64 \, ,
\eqlabel{gbb}
\end{equation}
while for the other values of $\b$ we consider, it dips below this bound.
As we observed in the thermodynamic analysis of the broken phase, see Fig.~\ref{figure2},
decreasing  $\b$ makes the broken phase (while still thermodynamically preferable)
closer and closer to the unbroken phase. Correspondingly, for smaller values of $\b$
the shear viscosity is closer to the universal result down to lower and lower temperatures.
However, for sufficiently low temperatures it drops even steeper.
It is technically challenging to perform our numerical analysis reliably
at temperatures lower than those reported; nonetheless, the data obtained suggests
that the holographic plasma  \eqref{ea1} does \emph{not} have any lower bound on the ratio of the
shear viscosity to the entropy ratio as one varies $\b$.

Finally, we note that for positive values of $\b$ the broken phase of the holographic plasma
\eqref{ea1} has a shear viscosity ratio exceeding the universal result; we have not studied this
parameter regime in detail.

\section{Causality of holographic superfluid plasma}

Consistency of a holographic plasma as a relativistic quantum field theory
requires that it does not propagate modes faster than the speed of light.
The dispersion relation of the linearized fluctuations in the plasma is identified
with the dispersion relation of the quasi-normal modes of a black hole in the
dual gravitational background.
There are three types of quasi-normal modes in gravitational geometries with
translationally invariant horizons \cite{ks}:
\nxt a scalar channel
(helicity-two graviton polarizations);
\nxt a shear channel
(helicity-one graviton polarizations);
\nxt a sound channel (helicity-zero graviton polarizations).\\
In the case of the GB plasma, the lower bound  on the shear viscosity
(the upper bound on $\lgb$ \eqref{b1}) comes from the scalar channel
quasi-normal modes \cite{rob1}. On the other hand the  upper bound on the shear viscosity
(the lower bound on $\lgb$ \eqref{b2}) comes from the sound channel
quasi-normal modes \cite{bm}. In our case, the study of the quasi-normal modes in the sound channel
%study of a sound channel quasi-normal modes
is the most difficult --- it requires understanding holographic
viscous hydrodynamics in the presence of Goldstone modes associated
with the spontaneous breaking of a global $U(1)$ symmetries. To our knowledge
such theory has not been developed yet\footnote{For a  step in
this direction see \cite{kovtun}.}. Instead, as in \cite{rob1},
we limit our discussion to the scalar channel quasi-normal modes.
We expect that analysis of the other channels will not change
our conclusions with regards to causality.

Our discussion here follows closely \cite{rob1}.
Due to their complexity, we omit most technical
details\footnote{Omitted expressions are available from the authors
upon request.}.
The quasi-normal equation for the scalar channel fluctuations for
the holographic plasma dual to \eqref{ea1} takes the form
\begin{equation}
Z''_{[scalar]}(r)+\calc^{(1)}_{scalar}\ Z'_{[scalar]}(r)+\calc^{(2)}_{scalar}\ Z_{[scalar]}(r)=0 \, .
\eqlabel{zs}
\end{equation}
Following \cite{rob1}, it is possible to introduce a new radial coordinate $y=y(r)$, with $y\to -\infty$
corresponding to the horizon and $y\to 0_-$ corresponding to the boundary,
and to rescale the radial profile as
\begin{equation}
Z_{[scalar]}=\frac{1}{\calg}\ \psi_{\rm [scalar]} \, ,
\end{equation}
such that \eqref{zs} can be brought into the form of an effective Schr\"odinger equation:
\begin{equation}
\begin{split}
&-\hbar^2\ \del_y^2\, \psi_{[\rm scalar]} +U_{[\rm scalar]}\
\psi_{[\rm scalar]}
=c_s^2\ \psi_{[\rm scalar]}\,,\qquad \hbar\equiv \frac {1}{\ka}\,,
\qquad c_s=\frac{\ww}{\ka}\\
&\qquad{\rm where}\ \ \ U_{[\rm scalar]}=U^0_{[\rm scalar]}+\hbar^2\
U^1_{[\rm scalar]}\,,\qquad \del_{c_s} U^0_{[\rm scalar]}=0\,,
\qquad \del_{c_s} U^1_{[\rm scalar]}\ne 0 \, .
\end{split}
\eqlabel{sscalar}
\end{equation}
Notice that in the limit $\ka\to \infty$ (or $\hbar\to0$), everywhere except in the tiny region $y\gtrsim
-\frac{1}{\ka}$ the dominant contribution to the effective potential $U_{scalar}$ comes from
$U^0_{scalar}$. Thus, in this limit it is a good enough approximation to take
\begin{equation}
\hbar^2\, U^1_{[\rm scalar]}=\begin{cases} 0 & \text{$y<0$\,,}
\\
+\infty &\text{$y\ge0$\,.}
\end{cases}
\eqlabel{u1scalar}
\end{equation}

As explained in \cite{v2,rob1}, the bound states of the resulting 1-dimensional quantum mechanical
problem \eqref{sscalar} with ``energy'' $c_s^2>1$ point to the presence of quasi-normal
modes in the plasma, propagating faster than the speed of light.
On the other hand, bound states with energy $c_s^2<1$ indicate the presence of
instabilities (tachyonic modes in the plasma in the limit $\ka\to\infty$).

\subsection{Causality of the symmetric phase}

\begin{figure}[t]
\begin{center}
\psfrag{r}{{$r$}}
\psfrag{us}{{$U_{[scalar]}$}}
  \includegraphics[width=4in]{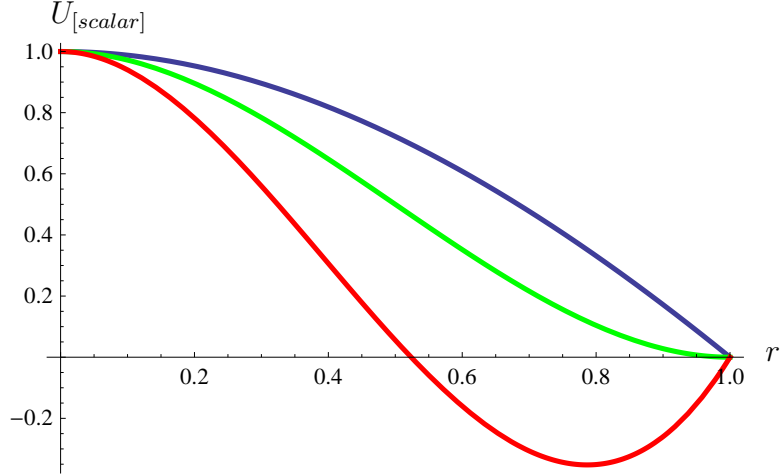}
\end{center}
  \caption{(Colour online) Effective scalar potential in the symmetric phase,
see \eqref{unbroken}. The blue, green and red curves correspond to, respectively, $\a_0=\{1,3,5\}$.
The regime $\a_0>3$ is unphysical, since it corresponds to negative
temperatures.}\label{figure4}
\end{figure}

The effective potential $U_{[scalar]}$ defined in \eqref{sscalar} and \eqref{u1scalar}
can be computed analytically in the symmetric phase, and is given by:
\begin{equation}
U_{[scalar]}=\begin{cases}
1-r^2\ \left(1+\frac 29 \a_0^2\right)+\frac29 \a_0^2\ r^3\,,\qquad &0<r\le 1 \; ,\\
+\infty\,,\qquad r\le 0 \; .
\end{cases}
\eqlabel{unbroken}
\end{equation}
Fig.~\ref{figure4} shows the potential $U_{[scalar]}$ above for select values of $\a_0$.
The blue, green and red curves correspond, respectively, to $\a_0=\{1,3,5\}$.
In the $\a_0=1,3$ plots, $U_{[scalar]}$ decreases monotonically between the boundary and the horizon,
and never develops a maximum in the intermediate region.
In all cases there are no bound states with energy $c_s^2>1$ -- there are no
superluminal quasi-normal modes and, as expected, the theory is causal.
Notice that for $\a_0>3$ the potential \eqref{unbroken} develops a negative energy minimum,
which implies the existence of negative energy bound states, and as
a result tachyonic (unstable) quasi-normal modes in the $\ka\to\infty$ limit.
However, this does not cause any problems: for $\a_0 >3$
the temperature of our plasma in the symmetric phase
\beq
T=\frac{z_0}{\pi} \left(1-\frac{\a_0^2}{9}\right)
\eeq
becomes negative, and the tachyons are therefore not physical.
%negative temperatures in plasma, see \eqref{tsym}.

\subsection{Causality of the broken phase}

\begin{figure}[t]
\begin{center}
\psfrag{r}{{$r$}}
\psfrag{us}{{$U_{[scalar]}$}}
  \includegraphics[width=3in]{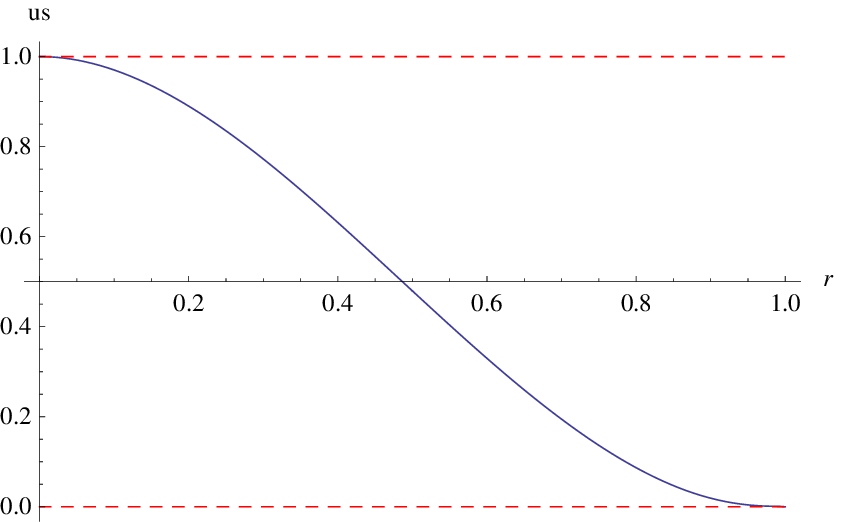}
  \includegraphics[width=3in]{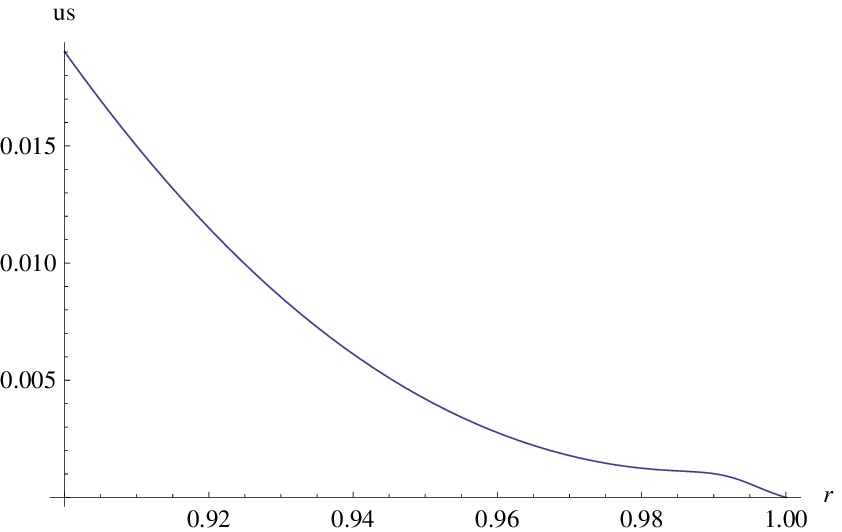}
\end{center}
  \caption{(Colour online) Effective potential $U_{[scalar]}$ in the symmetry broken phase
of the holographic plasma \eqref{ea1}
at $\b=-5$ and $\frac{2\pi T}{\mu}=0.01(9)$, corresponding
to $4\pi \frac{\eta}{s}=0.4(9)$ .
The dashed red lines correspond to $c_s^2=\{0,1\}$.
}
\label{figure5}
\end{figure}

In the symmetry-broken phase the effective scalar potential $U_{[scalar]}$ \eqref{sscalar}, \eqref{u1scalar}
can only be computed numerically.
A representative example of such computation is shown in
Fig.~\ref{figure5}. Here,  $\b=-5$ and $\frac{2\pi T}{\mu}=0.01(9)$, corresponding
to $4\pi \frac{\eta}{s}=0.4(9)$ (the low temperature endpoint of the
orange curve in Fig.~\ref{figure3}).  Notice that this potential does not support
bound states with energy $c_s^2>1$; neither does it support states with $c_s^2<0$.

We conclude that, at least in the scalar channel, the gauge theory plasma
holographically dual to the gravitational model \eqref{ea1} does not violate causality.
We also find that it does not contain any tachyonic modes in the $\ka\to\infty$ limit -- the theory appears
to be perfectly well-behaved over the entire range of parameters. Unlike the case of the GB plasma (\ref{gbg}), here the
self-consistency of the CFT doesn't place any constraints (whether from below or above)
on the size of the higher-derivative coupling.
While it is possible that the shear and sound channels might lead to additional
instabilities, previous studies \cite{bm,gc2} suggest that this should not be the case.

\section{Conclusion}

In this paper we have argued that microscopic constraints
(causality, positivity of energy, etc.), while important for the general consistency
of a plasma as a relativistic quantum field theory, are not necessarily responsible for
setting the lower bound on the ratio of shear viscosity to entropy density in the plasma.
The basic reason is that the hydrodynamic transport of the system is determined
by its infrared properties, which do not necessarily enter into the microcausality
analysis of the theory. To this end, we generalized the holographic model of ``GB plasma''
introduced in \cite{rob1} in such a way that the Gauss-Bonnet coupling of the former is
replaced with an (irrelevant) operator. Our holographic model, see \eqref{ea1},
undergoes a second order phase transition at low temperatures, where this operator
develops a vacuum expectation value. As a result, the effective Gauss-Bonnet coupling
in our model is nonzero in the broken phase  (which is necessary to generate
the non-universal ratio of shear viscosity to entropy density),  but being
identified with an irrelevant operator it does not effect the ultraviolet properties of the
model --- the dynamics at high energies is equivalent to that of holographic superconductors
\cite{gub1}.

We identified parameters in our model where the shear viscosity drops below the
causality bound\footnote{The current lowest bound on the ratio of the shear viscosity to the
entropy density in 4-dimensional plasma was reported in \cite{rob2}. It is not clear though whether
the model discussed there is a consistent relativistic QFT.}
\begin{equation}
\frac{\eta}{s}\ge \frac{1}{4\pi}\ \frac{16}{25}\,,
\end{equation}
determined in \cite{rob1}. It would certainly be interesting to identify the lowest bound in our model
--- however, this is not the main focus of this paper. It is clear that, whatever the
lowest bound (assuming it exists) on the shear viscosity ratio in holographic
plasma \eqref{ea1}, it does not affect its causal properties.
To complete the analysis one would need to study causality in the vector and the sound
channels of the plasma quasi-normal spectrum\footnote{The analysis of the sound quasinormal models
are most challenging and will be reported elsewhere.} \cite{bm}.
As we already stated, we do not believe that
such analysis would modify the physical picture presented here.

To summarize, the question of the bound on the ratio $\frac{\eta}{s}$ suggested by a
quasi-particle picture of the fluid, its very existence,
and the physics that determines it  remains open.

\section*{Acknowledgments}
We would like to thank  Ofer Aharony, Micha Berkooz, Ramy Brustein, Jim Liu,
Rob Myers and Aninda Sinha for interesting discussions.
A.B. would like to thank the Mitchell Institute for Fundamental Physics and Astronomy, the
Weizmann Institute for Science and the Aspen Center for Physics for hospitality during various stages of this project.
Research at Perimeter Institute is supported by the Government of Canada through Industry Canada and by the
Province of Ontario through the Ministry of Research \& Innovation.
A.B. gratefully acknowledges further support by an NSERC Discovery grant and support
through the Early Researcher Award program by the Province of Ontario.
The work of S.C. has been supported by the Cambridge-Mitchell Collaboration in Theoretical
Cosmology, and the Mitchell Family Foundation.

\end{document}